\documentstyle{article}
\begin{document}
\begin{center}  
\Large
The Broad Line Regions of NGC 4151\\
\normalsize
I.Cassidy \& D.J.Raine\\            
Astronomy Group\\
University of Leicester, Leicester UK\\
Revised 17 July 1995\\
Accepted for publication in Astron. \& Astrophys. 
\end{center}

\\
For figures send postal address or FAX number to jdr@leicester.ac.uk\\
\\
\begin{abstract}
We estimate the properties of clouds that form as a result of the interaction 
of the surface of a disc embedded in a supersonic outflow and heated by
external irradiation. 
We find two regions where short-lived clouds are injected into the 
outflow and accelerated by ram pressure. We indentify these with a broad line
region and an intermediate line region.
We compute the line strengths and profiles from a model of the cloud dynamics 
and compare the
results with the observations of the nearest Seyfert galaxy NGC 4151, 
which, in its various luminosity states, has provided a large amount of
emission line data.
We show that the model parameters for this galaxy are constrained by
just the CIV $\lambda 1549$ line profile in the 
high luminosity state of the nucleus. We review briefly 
some of the data on broad line ratios, profiles, profile variations and
transfer functions and
show how much (if not all) of this data can be accounted for in
the model. We also indicate how the X-ray data
can be fitted into this picture.
                                              
\end{abstract}                       
\section{Introduction}
We can distinguish various approaches to understanding the broad emission lines
from active galactic nuclei. In phenomenological models we attempt to obtain a
distribution of photoionised gas that will reproduce the main features of the
line ratios and profiles without regard to the dynamics of the gas or its 
relation to the system as a whole. Early investigations resulted in the now
universally accepted cloud picture, but were ambiguous with regard to the cloud
kinematics. More recent investigations of time lags between continuum and line
variations restrict the scale of the cloud distribution and point to  complex
kinematics and a
stratification in cloud properties with radial distance from the ionising
source. To accomodate these observations, phenomenological models must incorporate 
a range of
parameters or arbitrary functions (for example, Robinson 1995). 
An alternative approach is to try to model
the cloud system dynamically. Gravity, magnetic fields, 
ram pressure and radiation pressure have
all been considered for the acceleration of pre-existing clouds (Carroll \&
Kwan 1985, Emmering et al. 1991, Capriotti, Foltz \& Byard 1979, 
Shields 1978, Blumenthal \&
Mathews 1975, Mathews 1993). Two phase
equilibria and shocks have been considered for the formation process (Mathews
\& Doane 1990, Perry \& Dyson 1985). 
Models of
this sort have been used to address either individual galaxies or classes of
galaxies (Collin-Souffrin et al. 1988, Terlevich 1994). 

The present paper stems from a phenomenological model (Cassidy \& Raine, 1993)
in which
clouds are supposed to be injected from the surface of an accretion disc into
a supercritical, mildly relativistic, outflowing spherical wind where they 
are accelerated and destroyed. Although the
existence of such a  wind is an assumption for the formation of which there is no 
firm theoretical evidence, we showed that
this picture gives a good account of the large variety of line profile shapes that
occur in active nuclei and we were able to relate characteristic features of
the profile types to aspects of the physical model. 
This work was in turn based on a theoretical model
(Smith \& Raine 1984) of the
formation and acceleration  processes in a thermally driven wind from the disc
surface which we now know to be untenable for the broad lines (Sect.
2). Here we present a new physical picture of the interaction of an accretion
disc in an ambient wind in which the broad line clouds are formed as a result of 
Kelvin-Helmholtz instabilities at the interface. A further 
new feature of the model is
that the thermally driven outflow from the disc provides a region of
line emitting clouds (the intermediate line region, ILR) 
with properties intermediate between the true broad line
region (BLR) and the narrow line region (NLR). The theory supports the phenomenology
described in Cassidy and Raine. It enables us to go some way towards a detailed
description of the broad line emission from NGC 4151, one of the most enigmatic (and
hence theoretically challenging) of galaxies. In Fig.1 we give a schematic
diagram of the main features of the model.

The picture we present attempts to address seriously from a theoretical
viewpoint the wealth of detailed
observational data that point to a complex, stratified, emission line region. Our
model is similar to Emmering et al. (1991) except that they 
take magnetic forces to be partly
responsible for the injection and acceleration of cloud material from the
surface of the disc into the broad line region. It differs from biconical
models (Zheng, Binette \& Sulentic 1990) in that our cloud outflow fills a conical 
shell rather than the
interior of a (double) cone and our cloud motion inherits an azimuthal
component from the disc. Conversely, we differ from pure disc models (e.g.
Shields 1978) since in our case there is a radial component to the cloud velocity. 
In contrast to the two component model of 
Collin-Souffrin et al (1988), we obtain also the low ionisation lines
from the cloud component and not from disc emission. The two component model is
certainly not applicable to NGC 4151 in any obvious way since in this source the 
CIV $\lambda 1549$ and MgII $\lambda 2798$           
line profiles are similar to each other 
and both different from CIII] $\lambda 1909$. 

Clearly our
picture is not spherically symmetric so differs from, for example, 
Dyson and Perry (1985),
Mathews (1993), Terlevich et al (1994) and many others. Note that we do inject
short-lived
clouds into an outflowing wind but (in contrast to, for example, Dyson and
Perry) the cloud material does not originate in the wind. Nevertheless, we
agree with many authors in taking the components of the active nucleus to be 
a disc, a nuclear source of both radiation and particle flux and stars. We differ in
that we explore the interaction of the nuclear wind and the disc as the source
of emission line clouds. Of course, 
our nuclear wind must interact with the stars in the BLR as in the Dyson and
Perry picture; but we assume that, for NGC 4151 at least, cloud formation from shocks
induced in the wind by the present 
population of supernova remnants is too small to make a significant
contribution to the BLR (although it may make some contribution to the ILR). We
also neglect the contribution from bloated stars (Alexander \& Netzer, 1994)
and the stellar ionising continuum.

NGC 4151 exhibits changes of state that can be described as a sequence of
intermediate Seyfert types. Ulrich et al. (1991) distinguish between a 
high luminosity state (f$_{\lambda 1455} > 15 \times 10^{-14}$ erg cm$^{-2}$
s$^{-1}$ \AA$^{-1}$) and a low luminosity state (f$_{\lambda 1455} < 3 \times 
10^{-14}$ erg cm$^{-2}$ s$^{-1}$ \AA$^{-1}$). 
At its highest luminosity NGC 4151
shows features characteristic of Seyfert I nuclei with broad optical
and UV lines (up to 30 000 km s$^{-1}$ FWZI) and a high H$\beta$/[OIII] ratio
(Clavel et al. 1990). At its lowest luminosity the broad lines have all but
disappeared and the nucleus is described as a Seyfert 1.9. Changes between 
states take place on a time
scale of weeks to months, short even in comparison with other such intermediate 
Seyferts. However, the extreme luminosities occur very rarely. In section 3 we
shall show that the CIV line profile in the high luminosity state fixes the
parameters of the model. We consider also the likely effect of reductions of
luminosity using the theory of section 2 as a guide. In section 4 we explore
the extent to which the model might yield agreement with the extensive range of
observational data on this galaxy.

\section{The Theoretical Basis}                                  
\subsection{Introduction}
In this section we shall show that for an accretion disc embedded in a
supersonic outflow, and illuminated by an external radiation field, there are two
regions from which clouds are injected from the disc surface into the ambient 
outflow. The inner of these regions yields clouds with the properties of the
broad line gas (BLR) and the outer region can be identified with the intermediate
emission 
line region (ILR). We shall anticipate this identification and refer to the two
zones of the disc as the BLR and ILR.
The theory will be developed in general and applied to the source
NGC 4151; thus we shall write some numerical factors in equations in terms of
the parameters for this galaxy, even though, of course, this will require us to
assume the results of
the fit to the observations in sections 3.  

External illumination by X-rays raises the temperature of material at the
surface of the disc to the Compton temperature. At sufficiently large radii
this material is then hot enough to escape from the gravitational pull of the
black hole (Begelman \& McKee 1983). We have shown that if this material is
subject to the ram pressure from a nuclear wind then unstable cooling leads to
cloud formation (Smith \& Raine 1984). 
Our     
original proposal that this is 
the mechanism for the formation of 
broad line clouds is, however, untenable. For a typical Compton
temperature of the incident radiation
field, $ T_{c}=10^{7}$, mass loss from the disc 
can occur only at radii greater 
than $r_{c}=10^{19}m_{8}/T_{7}\:$ cm for a black hole of mass $10^{8}m_{8}M_{
\odot}$, where $T_{7}=10^{7}T_{c}$ 
Even if the Compton temperature of the radiation incident on the disc
differs from that in the line of sight to the observer (which is not the case
in our model of NGC 4151) the inferred radius of the supposed BLR would still
be larger than permitted by variability arguments. 
Furthermore, if the Compton temperature is
$10^{7}$K, the density in the clouds, at temperature $10^{4}$K, will be of order
$ 10^{8}$ cm$^{-3}$ (see equation (\ref{nlrdens})), too low to be compatible with
the typical inferred value of the BLR ionisation parameter of $\log U \sim -2$.      
In addition, in order to
obtain logarithmic line profiles in the line wings, we assumed 
that the clouds are long-lived. The longevity of the clouds
is inconsistent with acceleration by a supersonic wind (since the clouds loose
mass through the flanks on the order of a sound crossing time). Short-lived
clouds in this model cannot produce acceptable line profiles (Cassidy,
1994) nor provide the variety of non-logarithmic profiles now observed in other
sources (Robinson 1995). The requirements are relaxed somewhat if, as may be the
case in NGC 4151, the Compton temperature is as high as $10^{8}$K, but the
overall conclusions are unaffected. 

Nevertheless, this mass loss must occur, and, if the nuclear wind is present,
cloud formation is a consequence. We shall show that
these clouds must contribute to an intermediate line region. In the 
low luminosity
states of NGC 4151 the ILR dominates the line emission; from this we shall deduce
that these clouds may be sufficient to provide the main component of 
the ILR emission in this source.
 
We shall show further that within the Compton radius, $r_{c}$, the illuminated
surface of the disc in the presence of the nuclear outflow is Kelvin-Helmholtz
unstable. These instabilities, we argue, will grow and be entrained in the wind
as BLR clouds.

In Cassidy \& Raine (1993) we used a phenomenological picture based on the
injection of clouds from the surface of the disc by an unspecified mechanism 
to generate the broad line
region. This work gave a range of line profiles that could be
matched to the variety of profile
classes observed, but at the expense of a large number of adjustable
parameters since the physical basis of the model was not specified. 
In the present paper we take a different approach: 
we construct a theory of the cloud injection process which enables us to relate
some of the parameters of the system by self-consistency arguments. We then
apply the results to one of the most difficult examples, NGC 4151, for which we
can find a set of values for the remaining parameters from a single
observation of the CIV $\lambda$1549 line profile at its peak luminosity.
In particular we fix the angle of the disc to the line of sight.
In a later paper we shall show that NGC 4151 (in each of its various luminosity
states) can indeed be considered
a typical active nucleus viewed from the particular line of sight, just over
the top of the disc, at which the
nearside of the flattened distribution of BLR clouds appears directly in front of the
central continuum source.

Thus, to summarise, the active nucleus is supposed to contain 
a black hole, of mass $m_{8}$ in units of
$10^{8}M_{\odot}$; there is a disc, which is taken to be a source of material, but
about which we make no assumptions other than that it is opaque with non-zero
albedo; we require a central 
source of luminosity $L_{r}$, 
parameterised by $l_{r} = L_{r}/L_{E}$, where $L_{E}= 1.3\times 10^{46}m_{8}$ 
is the Eddington limit luminosity; and we have a mass outflow which carries a
flux of kinetic energy $L_{m}$. This matter flux, the Eddington luminosity and
the disc inclination
are the only true free
parameters in the theory, apart, that is, from dimensionless parameters of order
unity, behind which we hide our ignorance of some of the details of the physics. The
results depend significantly on 
only one of these dimensionless parameters, the disc albedo.
We also have to
make assumptions about the continuum spectrum in those bands where it is not
measured, but our results are not sensitive to this choice, 
except for some
dependence on the Compton temperature of the disc illumination (which is 
not necessarily the same as the observed Compton temperature). As noted below, 
we prefer a Compton temperature of the illuminating radiation closer to
$10^{8}$ K than $10^{7}$ K.
The wind must be sufficiently optically thin to
be compatible with the X-ray variability. We use the observations of NGC 4151 to
calibrate this constraint. The interactions in this system then lead to a scattering
of radiation in the wind on to the disc, a zone in which the disc surface is
Kelvin-Helmholtz unstable, which is the BLR, and a zone in which the disc
surface is the source of a thermally driven wind, which forms the ILR (Fig.1).
\subsection{Wind Properties}                                                    
The supersonic nuclear wind is taken to be spherical with a 
constant outflow velocity 
$v_{w} = 10^{10} v_{10}$ cm s$^{-1}$ and to carry a power
$L_{m}$ in bulk motion.
The wind density, $n_{w}$,
is then given by
\begin{equation}
n_{w} = 4 \times 10^{6}(L_{m}/L_{E})(m_{8}/0.3)r_{16}^{-2} v_{10}^{-3} 
{\rm \: cm}^{-3},
\end{equation}
and the ram pressure is
\begin{equation}
P_{m} = 7 \times 10^{2} (L_{m}/L_{E})(m_{8}/0.3)r_{16}^{-2}v_{10}^{-1} 
{\rm \: erg \: cm}^{-3}.
\end{equation}
The wind is fully ionized by the radiation field and heated to the Compton
temperature out to a radius $r_{ad}=3\times 10^{17} l_{r} m_{8}
v_{10}^{-1}$
cm. Beyond this adiabatic cooling will start to become
important but the wind temperature will not drop substantially until
$r=10^{3}r_{ad}$ (Smith and Raine 1985).
The Mach number of the outflow is of order $100 v_{10} T_{8}^{-1/2}$ and
the wind is typically hypersonic. 
The optical depth to electron scattering is
\begin{equation}
\tau_{w} = 2 \times 10^{-2}(m_{8}/0.3)(L_{m}/L_{E})r_{16}^{-1} v_{10}^{-3}.
\end{equation}
So, for NGC 4151, provided $L_{m}/L_{E} \leq 5$, we have $\tau_{w}$ $\sim$ 1 
at $r = 10^{15}$ cm which is consistent with
large scale X-ray variability on times of order $10^{5}$ s. (Yaqoob et al.
1991, 1992). 
Then, consistently, $v_{w}$ = $10^{10}$ cm s$^{-1}  \gg v_{esc}$ at
$10^{15}$ cm (since 
$v_{esc}$ = 10$^{10}((m_{8}/0.3)/r_{14})^{\frac{1}{2}}$ cm s$^{-1}$). We
cannot reduce the outflow velocity significantly below $v_{10}=1$ for a
high value of $L_{m}/L_{E}$, as will be required to fit the high luminosity
line profiles below. The
nuclear outflow must therefore be mildly relativistic for consistency.
                                                                        
\subsection{Properties of the disc}
The radiation falling on the disc will significantly affect the surface if the
energy falling on the disc is greater than that produced locally by viscosity
in the disc. 
In particular the surface layers will be heated to the Compton
temperature. Note that under these circumstances the disc may still have an
optically thick, cool, geometrically thin core (Mobasher \& Raine 1990) at this
radius.
Suppose the disc is illuminated directly (rather than by scattered
radiation). Let the surface be given by $z=z(r)$. Then the luminosity
intercepted by the disc between radii $r$ and $r+{\rm d}r$ is
proportional to $2\pi r {\rm d}z/r^{2}$. Let the flux falling on the disc
be $(\phi L_{r}/4\pi r^{2}){\rm d}z/{\rm d}r$ between radii $r_{1}$ 
(where $\tau_{w}=1$) and $r_{max}$. (This defines $\phi$.) Taking $z(r)$ to be
given by a scale height in the hot surface layers of the disc 
when external illumination dominates gives
$z(r)=\times 10^{15}T_{8}^{1/2}r_{16}^{3/2}(m_{8}/0.3)^{-1/2}$ cm. 
For the local dissipation in
the disc it is reasonable to assume a steady state. But the irradiation from
the central source may change on short time scales. 
In terms of a time-averaged
luminosity parameter $\bar{l_{r}}$, we therefore have $L_{r}=
0.1(\bar{l_{r}}/l_{r})\dot{M}c^{2}$ for an accretion efficiency of
0.1 and an accretion rate $\dot{M}.$
This gives the radius at which external illumination equals the local rate of
viscous energy generation as 
\begin{equation}                    
r_{in} \sim 10^{16}(\frac{m_{8}}{0.3})^{2/3} (\frac{0.1}{\phi})^{2/3} 
                T_{8}^{-1/3} (\frac{\bar{l}_{r}}{l_{r}})^{2/3}
                   \; {\rm cm}.                                          
\end{equation}                                 
This therefore gives an estimate of 
the inner radius of the BLR which consequently ranges between
about $10^{16}$ cm and $5\times 10^{16}$ cm as $l_{r}$ changes by a factor
10.
The outer radius of the BLR depends on the flaring of the disc in the presence
of the wind and is
calculated in section (2.7) below.  

The surface temperature of the disc is the Compton temperature. Beyond a radius
$r_{c}=3 \times 10^{17}(m_{8}/0.3)T_{8}^{-1}$ cm this exceeds the 
temperature at which material is
gravitationally bound (Begelman et al 1983). The material escaping from the disc
has been raised to a pressure $P=P_{rad}/\Xi'_{c,min}$ where $\Xi$ is the ionisation
parameter
$\Xi=P_{rad}/P_{gas}$ and $\Xi'_{c,min}$ is its value at which the gas can just
remain in the hot phase. The escaping material 
therefore has a density $\rho=P/c_{s}^{2}.$ Exposed to the ram pressure of the
nuclear wind this material cools under constant pressure to $10^{4}$K. 
Using $\Xi'_{c,min}=3$ as a typical value for a Seyfert spectrum, 
the density of the clouds becomes
\begin{equation}
n_{ILR}=10^{8} \frac{(l_{r}/.1) (\phi/.1)                       
T_{8}^{2}}{(r/r_{c})^{2}(m_{8}/0.3)} {\rm cm}^{-3}.\label{nlrdens}
\end{equation}
These clouds constitute the ILR.
Begelman et al. (1983) find that there is a fully developped disc wind
extending from $r_{c}$ to a radius
$r_{iso}=3(\bar{l}_{r}/0.1)T_{8}^{-1/2}r_{c}$ within which material is
rapidly heated to $T_{c}$.  
Thus the ILR extends from within $3
\times 10^{17}$ cm to beyond $10^{18}$ cm.

Most models of accretion discs are optically thick to the continuum in the BLR,
but not usually in the ILR (Collin-Souffrin 1987, Mobasher \& Raine 1990). 
Therefore  the disc obscures any
emitting material in the BLR on the far side from the observer, 
but not in the ILR. In the BLR we
expect the albedo of the disc to be significant at least for resonance lines.
Thus, we shall have to take account of reflection from the disc surface.

\subsection{Disc-wind interaction in the BLR}                                  
Consider now the interaction between the disc and the nuclear wind in the BLR
region. 
Suppose, contrary to our discussion above, that the
disc surface were to remain cold and therefore essentially
unexposed to the ram pressure of the
wind. Then the pressure in the disc is balanced by the thermal pressure,
$n_{w}kT_{c}$ in the
wind. If a fraction $\phi$ of the radiation field falls on the
disc, the ionization parameter (Begelman, McKee \& Shields 1983) is therefore  
\begin{equation}
\Xi = \frac{P_{rad}}{P_{gas}} = \frac{(\frac{\phi L}{4 \pi r^{2}
c})}{(n_{w}kT_{c})} 
 = 2 \times 10^{3} \frac{\phi L v_{10}^{3}}
{L_{m} T_{8}} \label{xi}
\end{equation}   
for an inverse Compton temperature $T_{c} = 10^{8} T_{8}$ K. By running the CLOUDY
photoionisation code for an average quasar spectrum  (Mathews and Ferland
1987), a power law with an energy index of -0.9, and a power law with the
canonical slope of -0.7,
we find that the material must be hot,
with temperatures of $1.5\times 10^{7}$, $1.0\times 10^{8}$ and $1.6 \times 10^{8}$
at values of $\Xi$ greater than of order $10^{2}$, 10 and 1 respectively. 
The value of
$\Xi$ from equation (\ref{xi}) is therefore much greater 
than the maximum $\Xi '_{c,max}$ for the cold phase. Thus, contrary to our
assumption (and in agreement with section (2.3)) the disc surface cannot remain
cold. But, also, the disc
surface cannot be hot and be exposed to the {\em full} ram pressure of the wind since
in that case 
\begin{equation}
\Xi = \frac{P_{rad}}{P_{m}} = (\frac{\phi L}{L_{m}})(\frac{v_{w}}{c}),
\end{equation}
smaller than a critical value 
$\Xi_{c}=\Xi_{h,min}'$, the minimum for the hot phase. 
So, we can assume that the disc surface adjusts itself so that the disc
is exposed to a pressure $f P_{m}$ just
insufficient to maintain the cold phase:
\begin{equation}
f = \phi l_{r}(\frac{L_{E}}{L_{m}}) (\frac{v_{w}}{c}) \frac{1}{\Xi_{c}}.
\end{equation}
If, for example, $L_{m}/L_{E} \leq 5$  and $l_{r}\phi$ =
0.01, we have $f \geq 2 \times
10^{-4}(3/\Xi_{c})$. 

From this we can deduce the density in the surface of the disc, $n_{d}$, 
since the disc
pressure must balance $fP_{m}$ at the Compton temperature. We obtain
\begin{eqnarray}        
n_{d}&=&10^{10}f(L_{m}/L_{E})r_{16}^{-2}v_{10}^{-1}T_{8}^{-1} \:{\rm \:
                 cm}^{-3} \nonumber\\
     &=&1.6 \times 10^{8}(\phi/0.1)(l_{r}/0.1)(m_{8}/0.3)
               \Xi_{c}^{-1}r_{16}^{-2}T_{8}^{-1}
              \:{\rm cm}^{-3}
\label{disc_n}
\end{eqnarray}

Next we show that this arrangement is Kelvin-Helmholtz unstable at 
the disc-wind interface. Rotation and infall of the disc material can be
neglected since the rotational and infall speeds are small
compared with the wind speed. Thus, for $n_{d} \gg n_{w}$
the instability operates on wavenumbers (Chandrasekhar 1961)
\begin{equation}
k > \frac{gn_{d}}{n_{w}v_{w}^{2}},
\end{equation}
where $g$ is the
gravity at the surface. Suppose the surface is at height $z$ = $\zeta H$, where
$H$ is the pressure scale height at the temperature of the disc atmosphere,
$10^{8}T_{8}$ K. Then, the vertical component of the acceleration due
to gravity is, approximately for $z \ll r,$
\begin{equation}
g = \frac{GM_{bh}z}{r^{3}} = 6.4 (m_{8}/0.3) T_{8}^{\frac{1}{2}} 
       r_{16}^{- \frac{3}{2}}
          \zeta \; {\rm cm \: s}^{-2}.           \label{g}
\end{equation}    

The height $\zeta$ is determined by pressure balance in the disc. For
any disc model it is unlikely that $\zeta \gg 1$ (because the pressure
falls off rapidly with height as exp(-$\zeta ^{2}$)), or $\ll 1$ (because the
external pressure does not dominate the internal disc pressure). It is 
therefore sufficient to take $\zeta = 1.$

Using equation (\ref{disc_n}) for the density in the disc, we obtain
\begin{equation}
k > k_{min} = 2.6 \times 10^{-16}
(m_{8}/0.3)(L_{E}/L_{m})(\phi/0.1)(l_{r}/0.1)\zeta
v_{10}T_{8}^{-1/2}r_{16}^{-3/2}\Xi^{-1} \: {\rm cm}^{-1},
\end{equation}
and potentially large columns
\begin{equation}
n \lambda_{KH}^{max} = 4 \times 10^{26} (L_{m}/L_{E})(m_{8}/0.3)r_{16}^{-1/2} 
v_{10}^{-1} T_{8}^{-1/2}\zeta^{-1} \; {\rm cm}^{-2}, 
\end{equation}                                                   
where 
\begin{equation}
\lambda_{KH}^{max}=2\pi/k_{min}= 2\times 10^{18} (0.01/\phi l_{r})(0.3/m_{8})(L_{m}/L_{E})    
                \Xi_{c} v_{10}^{-1} \zeta^{-1} r_{16}^{-3/2}T_{8} \: {\rm cm}
\end{equation}
is the length scale of the growing perturbations. 
Note that at the inner radius
this is larger than the BLR, so this maximum, which is calculated assuming a
uniform system, cannot be attained. The outer radius of
the BLR will turn out to be of order $10^{17}$ cm, (section 2.7) so
$\lambda_{KH}^{max}(r_{out})\sim r_{out}$ i.e.the maximum scale of the growing mode 
is of the order of the size of the BLR. However, the growth time decreases for
smaller wavelengths (and is infinite for the
largest mode (equation (17)) so  
the disc surface will develop large rising rings of hot gas on various scales
within the BLR. 
This gas is  exposed to
an increased pressure from the impact of the nuclear wind at which the material
cannot remain hot. 

The main contribution to cooling above about $10^{6}$K is thermal
bremsstrahlung, so the cooling timescale in the disc surface is
\begin{equation}
t_{c} \sim \frac{nkT}{10^{-27}n^{2}T^{1/2}} \sim 10^{7} (0.1/\phi)(0.1/l_{r})
           (0.3/m_{8})T_{8}^{3/2} \Xi_{c} r_{16}^{2} \: {\rm sec}.
\end{equation}
This is to be compared with the sound-crossing time, $t_{s}$. For material of column
density, $N_{col}$, we have $t_{s}=N_{col}/nc_{s}$
and hence
\begin{equation}
t_{s}/t_{c} \sim (N_{col}/10^{23})T_{8}^{-1},
\end{equation}
independent of any parameters of the system.
Thus, the hot bubble can cool coherently only in a relatively 
small column of density about $10^{23}$ cm$^{-2}$.  Cooling to $10^{4}$ K
will tend to be one-dimensional producing a surface layer of this column
density. The growth time of the instability is
\begin{equation}
t_{g} = (\frac{k^{2} n_{w} v_{w}^{2}}{n_{d}})^{- \frac{1}{2}} (\frac{
k_{min}}{k} - 1 )^{- \frac{1}{2}}.
\end{equation}
which, for scales $\lambda_{KH} \ll \lambda_{KH}^{max}$ depends only on column
density so is unaffected by cooling. Perturbations on the cold surface grow on
a timescale of order $10^{-4}t_{s}$. Thus, the surface will tend to ripple as
it cools into clouds of number density, assuming cooling at constant pressure,
\begin{equation}   
       n_{c}= \frac{3 \times      
10^{12}}{r_{16}^{2}}(\frac{\phi}{0.1})(\frac{l_{r}}{0.1})
(\frac{1}{T_{8}\Xi_{c}}) \; {\rm cm}^{-3}.
\label{n_c}                                                    
\end{equation}
Where the timescale for cloud formation, $t_{c}$, 
exceeds the variability timescale the
factor of $l_{r}$ in this equation should be replaced by its time averaged
value $\bar{l_{r}}$. But for most, if not all, of the BLR we can ignore
changes in cloud density. These clouds will be entrained into the nuclear wind. 
                                 
We can now summarise how the instabilities develop. Hot rings grow on
the disc surface and  start to rise into the wind. The
material at the surface of the rising bubble is exposed to a higher pressure and
a shell of column of $N_{c}=10^{23}$ cm$^{-2}$ starts to cool towards $10^{4}$K. 
The cool material is still
Kelvin-Helmholtz unstable on scales less than $N_{c}$ 
so cooling does not suppress the growth but will tend to break the rings up
into clouds.. Thus, the effect
of the instability is to inject a spray of clouds of column density
$10^{23}$ cm$^{-2}$  into the nuclear wind. Let $\lambda_{c}=N_{c}/n_{d}$, the
thickness of the shell of disc material that cools in a sound crossing time.  
The efficiency with
which the hot bubble is injected as cold clouds depends on the ratio
$\lambda_{c}/\lambda_{KH}$, since this is the fraction of material in the
bubble that cools to form the surface shell.  Thus the number of clouds
injected as a function of radius is:
\begin{equation}
\lambda_{c}/\lambda_{KH}^{max}=3 \times 10^{-4}
(L_{E}/L_{m})(0.3/m_{8})v_{10}T_{8}^{1/2} \zeta r_{16}^{1/2}.
\end{equation}
The important point is that this increases with radius as $r^{1/2}$ thereby
weighting the outer clouds.

The density of the cold material at $10^{4}$K is
a factor $10^{4}T_{8}$ greater than the surface density of the hot disc
(equation (\ref{n_c})). Such densities (of order $ 3\times 10^{12}$ cm$^{-3}$ for the
innermost clouds in NGC 4151, falling to $3\times 10^{10}$ cm$^{-3}$ in the
outer BLR) are much higher than usually attributed to the
BLR. We shall find such densities to be precisely those required for agreement
with the CIV/CIII] line ratios and profiles.

\subsection{Cloud injection}
We argue next that these
cool regions are injected into the nuclear wind with some velocity
$v_{i}$ and accelerated to some velocity $v_{c}$. Consider an elevated
region of surface wave. The growing protuberance provides an obstacle
around which the wind must flow. While the face of the perturbation
is exposed to some fraction  of the ram pressure, 
the top surface will be subject to a smaller
pressure closer to the thermal pressure, $P_{w}$, in the wind. Since the
disc material is on average at $ f P_{m} \gg P_{w}$ the incipient hot
bubble, area $A$, is accelerated upwards to a velocity $v_{i}$ given by
\begin{equation}
f P_{m}A= (A \lambda_{KH}  n_{d} m_{p})(
\frac{v_{i}^{2}}{\lambda_{KH}}) .
\end{equation}
This gives $v_{i}^{2} = P_{d}/m_{p}n_{d}$ or $v_{i} = c_{d}$, the sound
speed in the hot disc surface. The sprays of cold bubbles are injected 
with velocities normal to the disc
of order $10^{8}T_{8}$ cm s$^{-1}$. Note that, even if the Compton temperature
of the disc surface
is of order of the observed average of $10^{7}$K, mechanical heating of 
the atmosphere may raise the temperature above $T_{8} \sim .1$ so injection
velocities above $3 \times 10^{7}$ cm s$^{-1}$ may be possible (King \& Czerny
1989). (In fact we shall take
$T_{8} \sim 1$ for NGC 4151 on other grounds; see sect.3)
The clouds will also inherit a rotational velocity from the disc. The disc
surface must flare to a scale height at the Compton temperature at the outer
edge of the BLR  (ie $H/r \sim 0.5$) so the rotational velocity will not be
Keplerian. However, since $H/r < 1$ throughout the BLR and $H/r < 0.1 $ for
much of the region, we shall take the azimuthal component of the cloud velocity
to be Keplerian on injection. In fact, the drag on the cloud as it moves across
the radially flowing wind reduces the rotational component $v_{\phi}$ significantly in
a time $(c_{s}/v_{\phi})t_{s}$, where $t_{s}$ is here the sound crossing time
in the cloud. Since the azimuthal velocity in the disc is supersonic, the
rotational component is strongly attenuated. So although it cannot be neglected
in forming the line profiles (since it is largest in the initial part of their
motion where the clouds spend the most time) departures from Kepler velocities
are unlikely to be of crucial significance. We therefore assume the clouds
inherit a Keplerian velocity from the disc on injection. 
The injected clouds will be subject to the full ram pressure of the wind.
This will act to compress and accelerate the clouds but also to destroy
them. We expect the cloud survival time to be of order a sound crossing
time in the uncompressed cloud. Thus, approximately, for the  cloud
acceleration, we have (ignoring mass loss)
\begin{equation}
\rho_{c} A  \lambda_{c} \frac{dv}{dt} = \rho_{w} v_{w}^{2} A
\end{equation}
or, taking $t = t_{s} = \lambda_{c}/c_{s}$ ($c_{s}$ is the sound speed in
the cold cloud),
\begin{equation}
v_{c} = c_{s} (\frac{v_{w}^{2}}{c_{s}^{2}})(\frac{\rho_{w}}{\rho_{c}}) =
f^{-1} c_{s},
\end{equation}
or
\begin{equation}
      v_{c}= 3 \times 10^{8}
\frac{\Xi_{c}}{(\phi/.1)(l_{r}/.1)v_{10}}(\frac{L_{m}}{L_{E}}) \: {\rm
cm}^{-1} \label{vc}
\end{equation}
using equation (8).
For $f \sim 5\times 10^{-4}$ we obtain speeds up to $2 \times 10^{9}$
cm s$^{-1}$. However, on the one hand, 
this estimate takes no account of  mass loss from the
clouds over their lifetime, whereas in the numerical computation of the cloud
trajectories this is incorporated through a simple model of cloud disruption
(see sect. 2.6).  This estimate, of the  cloud velocities should, as a result, 
be increased by a factor of, approximately, the ratio of the initial and
final column densities of the cloud, $(N(0)/N(t_{s}))$. On the other hand, the
numerical computation takes account of gravity and the full three dimensional
geometry with the overall result that the maximum final velocity 
(hence the line half-width at zero intensity) is of order $
10^{9}$ cm s$^{-1}$.

Notice that, other things being equal, the cloud velocity (equation (\ref{vc}))
appears to increase
as the luminosity in the radiation field, $l_{r}$, goes down. 
This is the converse of
what is observed for NGC 4151. However, we do not expect the power in the wind
to be independent of that in the radiation field. To produce  
the observed reduction in line width at low luminosity 
exactly in this theory we must require that the momentum flux in the wind,  
$L_{m}/v_{10}$, be proportional to $L_{r}^{q}$ with $q\sim 2$. 
This would be the case, for example, for a pair
driven wind.
We obtain qualitative agreement provided $P_{m}$
decreases with $L$.

\subsection{Cloud destruction}

Since the clouds are accelerated in a supersonic wind they are subject to
disruption by mass loss through the flanks, where the wind pressure is reduced
relative to the cloud face as the wind material accelerates round the cloud.
The results are not too sensitive to the manner in which the clouds break up,
provided that the result of evolution is to reduce the column density, 
so we take a simple model. We choose for the mass of the cloud 
$M(t)=M(0)(1+t/t_{s})^{-2}$ where
$t_{s}$ is the sound crossing time in the cloud. Thus, approximately 90\% of
the
cloud is lost in two sound-crossing times. The area of the cloud facing into
the wind and the matter density in the cloud are assumed to
be constant while the column density decreases with the mass. 

We can give some justification for this model as follows. 
Mass flow through the flanks of a cloud at the sound speed leads, after a time
$t$, to an increase in area facing the source to
$\pi r^{2}=\pi r_{0}^{2}(1+c_{s}t/r_{0})^{2}$, where $r_{0}$ is the initial
radius of the face of the cloud. This escaping material will be heated to the
Compton temperature and entrained in the wind. It is therefore lost from the
cloud. Thus the mass of the cloud is reduced by a factor $(1+t/t_{s})^{2}$
while the area
of cold cloud material remains constant.
Mass must flow through the cloud to supply the mass loss. Assume that 
the density in the cloud remains constant. (This would be a reasonable
approximation for a flattened  cloud
evolving  in pressure equilibrium, which is probably not too far from being the
case.) Consequently the column too decreases by this same factor.

\subsection{Outer limit of the BLR}
There is an outer limit to the cloud injection region set by the
extent to which the disc surface flares to intercept the momentum flux in
the wind.  The disc pressure must be provided by new wind material further from
the disc midplane as $r$
increases, because the cloud acceleration at smaller $r$ 
extracts momentum from the wind
close to 
the disc surface. At the outer edge of the BLR clouds rise to a height
\begin{equation}
z_{out} \sim t_{s}v_{i} \sim 10^{19} \frac{T_{8}\Xi
r_{16}^{2}}{(\phi/.1)(l_{r}/.1)} \; {\rm cm} \gg H(r_{out}).
\end{equation}
The momentum flux in the wind in a wedge of azimuthal extent
$\Delta \phi$ and altitude $\Delta \theta = z_{out}/r_{out}$ is, approximately
\begin{equation}
\Delta \phi \int_{0}^{\Delta \theta} sin \theta d \theta \: v_{w}^{2}
cos \theta \rho_{w} r^{2} \sim 10^{36}
\frac{r_{16}(L_{m}/L_{E})T_{8}\Xi_{c}}{(\phi/.1)(l_{r}/.1)}\Delta \phi.
\end{equation}
The maximum momentum flux in the clouds is 
\begin{equation}
\Delta \phi \dot{M}_{max} v_{c}= 2 \times 10^{36} \log r_{16}
[\frac{(l_{r}/.1)(L_{m}/L_{E})}{(\phi/.1)T_{8} \Xi_{c} v_{10}}]^{1/2}.
\end{equation}
We obtain equality when $r_{16} \sim 10$.
Thus, for the maximum cloud injection rate, the outer
radius of the BLR in NGC 4151 is at $10^{17}$ cm.

We have seen that the disc must flare to a scale height at the Compton
temperature at the outer limit of the BLR. For NGC 4151 ($m_{8}=0.3$) the scale
height is  $H = 1.5 \times 10^{15} T_{8}^{\frac{1}{2}}
r_{16}^{\frac{3}{2}}$ or $ 4.5 \times 10^{16}$ cm at $r_{out}=10^{17}$cm  
if $T_{8}=1$. So the disc flares to about .45 radians or about 30$^{o}$ across
the BLR. 
Beyond the BLR is a region in which the hot
disc atmosphere rises with $z \sim H \propto r^{3/2}$ until this material
escapes in a wind beyond $3\times 10^{17}$ cm. 

\subsection{The ILR}
The mass flow into the ILR from the thermally driven disc wind can be estimated
by assuming the heated material flows out at the sound speed; thus
\begin{equation}
{\rm d}\dot{M} = n_{d}c_{s}m_{p} (2\pi r {\rm d}r).
\end{equation}
The density in the disc surface $n_{d}$ is $10^{-4}n_{ILR}$ (equation (5))and
$c_{s}=10^{8}T_{8}^{1/2}$ cm s$^{-1}$. Assuming, as shown in Smith and Raine (1982),
that the material cools into clouds of
column density $\sim 10^{23}T_{8}$ cm$^{-3}$, we obtain for the sound 
crossing time in a cloud, $t_{s}$,
\begin{equation}
t_{s}=10^{9}(r/r{_c})^{2}(m_{8}/0.3)(0.1/l_{r})(0.1/\phi)T_{8}^{-1} \; {\rm s}.
\end{equation}
If the clouds survive for of the
order of a sound crossing time only, we obtain
\begin{equation}
dM=t_{s}d\dot{M}=2 \pi  10^{-3}T_{8}^{3/2}r{\rm d}r, 
\end{equation}
and hence, for the mass in clouds in the ILR,
\begin{equation}
M_{ILR} = 3 (\l_{r}/0.1)^{2}(m_{8}/0.3)^{2}T_{8}^{-3/2} M_{\odot}
\end{equation}
between the inner and outer radii given in section (2.3).
We shall show this is sufficient to provide the constant narrow component of
the BLR emission. For reference, the covering factor of the clouds (assumed
spherical), is of order $.01 T_{8}^{1/2}\log(3(l_{r}/0.1)T_{8}^{-1/2})$ and the
number of clouds is about $10^{3} (l_{r}
\phi/0.01)^{2}T_{8}^{1/2}.$ These estimates all lower limits since they 
assume the clouds survive only a sound crossing time.

\section{The model profiles}

The key to the comparison of our model with observations 
is the fit to the high state profile of the CIV line.
In our earlier paper (Cassidy and
Raine, 1993) we presented such a profile fit 
as an example of a Df class profile. This is reproduced in Fig. 2 here.

The requirement of sufficient cloud acceleration 
fixes $L_{m}/L_{E}$ (equation (\ref{vc})). 
We need
$L_{m}/L_{E} \sim 5$, an unexpectedly high value, since it implies that, in the
high state, most of
the energy output from the central source is in the nuclear wind. If we require
$L_{m}/v_{10} \propto l_{r}^{2}$ (section 2.5) then $L_{m}/L_{E} \ll 1$
in the low state and there is no overall problem in hiding the wind energy.
The profile fit,
given the column density,
fixes the inner radius at $r_{in}=10^{16}$cm and hence $m_{8} \sim 0.3$
(equation (4)). This, of course, gives us $L_{E}$ and hence
$l_{r}=L/L_{E}$ from observational estimates of the bolometric luminosity, 
$L$. We use $L/L_{X} = 10$ to estimate $L$ from the X-ray
luminosity $L_{X}$ (sect. 4.2) giving $l_{r}$ in the range
.1 to .01. In fact, with the radial extent of the BLR {\em fixed}, rather than
allowed to vary according to equation (4),  order of magnitude
changes in $l_{r}$ affect the line profile of CIV 
only through small changes in the core. Our chosen range of $l_{r}$ does,
howewer, allow a close match to the line ratios (sect. 4.3) and is consistent
with our estimate of the numerical factor in equation (4). Finally,
the injection velocity $v_{i}$ is fixed by the separation between the blue
peak and the central peak. 

Fig. 2 corresponds to the following choice of parameters:
black hole mass $m_{8}=0.3$, corresponding to $l_{r}(max)=0.1$;  
flux of energy in the nuclear wind  $L_{m}/L_{E} = 5$;
normal component of the cloud injection velocity 
$v_{i}=2 \times 10^{8}$ cm s$^{-1}$.
In addition, only clouds with column densities
corresponding to $10^{23}$ cm$^{-3}$ at the inner radius are considered.
                           
The continuum has been taken as the Mathews and Ferland spectrum for this
figure, although a power law gives essentially the same result.
The cloud injection velocity is consistent with a Compton temperature rather
higher than that for a Mathews and Ferland spectrum, but might be appropriate
for a power law with no EUV excess (Sect. 1). In any case much of the
radiation incident on the disc surface will be that emerging from the backs of
the clouds
forming close to the surface and will consequently be in the form of hard X-rays
with a raised Compton temperature. We therefore believe the required high injection
velocity to be reasonable.

In our
previous work the fit to the CIV line was made with the inner and outer radii
of the BLR, the cloud density and the disc flaring 
chosen freely. These are now, in principle, derived
quantities. We have allowed small factors between the order of magnitude
estimates of these quantities we have made for the model in section 2  
and the values chosen for the figure. We have used $r_{in} = 1.4 \times
10^{16}$ cm and $r_{out}=1.0 \times 10^{17}$ cm for the inner and outer radii
of the BLR, respectively, and $n = 3.6 \times 10^{12}$
cm$^{-3}$ for the density at $r_{in}$. The disc flares by 30$^{o}$ (sect. 2.7).

The best fit for the line of sight, given the parameter values above, is
58$^{o}$ to the disc normal. This fit we have judged by eye: it can be seen from
figure 2 that
the smooth model curve cannot fit the small scale detail of the profile, a
problem which is apparent in the $\chi^{2}$-values. In judging our claim
that the model does fit this observation one should bear in mind the
difficulties that all models have in reproducing even the broad features of the
profiles for this
particular galaxy. We should not expect the sort of fine detail in figure 2 to 
emerge here; we might expect it to be associated with the
fact that clouds are injected into the wind in bunches.
 
Our line of sight passes close to the disc surface as
required if the clouds
are to partially cover the X-ray source. In addition, the light curves 
are most simply explained if
there is emitting cloud material in the line of sight. The absence of a UV bump
in this galaxy might be accounted for by sufficiently 
anisotropic emission of the UV, 
although it is difficult to see how this might
be achieved physically.

Note that the double peakiness of the computed line in Fig 2 is a result
of a complex emission profile, and does not involve absorption.
However the dip between the peaks is shallower in the
model than in the observed line and almost absent if the disc albedo is
non-zero (Fig 2b).
This suggests that {\em some} true absorption must be involved in this feature.
Such absorption may arise if the inner CIV emitting clouds are obscured
by clouds further out. This is possible in NGC 4151 because of the
special angle to the line of sight. 
The variability in the depth and position of the absorption is also
in principle compatible with this picture.
The details do not affect the parameter values we obtain from the profile
fitting and
will be discussed elsewhere. We conclude that the observed CIV line
profile constrains the model parameters quite tightly and that the predicted
values of derived quantities provide significant
tests of the self-consistency of the model.

The disc itself is assumed to obscure either partially or totally
the line emission from clouds on the bottom side. (In section 4.6 we shall
argue in favour of total obscuration and a non-zero albedo, 
but we need to investigate the
properties of the partially obscured model to do this.) Without any obscuration
of clouds below the disc 
the line would be triple-peaked. In the case of partial obscuration
the line emission from each 
cloud is assumed to be blocked wherever there is
material moving in the disc along the line of sight with a relative 
velocity that is within a thermal width of the line centre.
What appears as the red peak in the observed line is therefore the central peak
of the unobscured profile. This will be relevant to the behaviour of the two
peaks when we consider variability in section 4. Finally, in section 4
we shall be led to consider a component reflected from the disc surface. 
In figure 2 we show the fit for both a partially obscuring disc with an albedo
of zero and for
a totally obscuring disc with an albedo
of 0.3. (In the latter case the disc is, of course, taken to be optically thick 
out to large radii, so the optically thin and partially obscuring regions
in Fig. 1 are absent.) 

\section{Comparison With Observations}

Detailed studies of NGC 4151 have been prosecuted now for about 25 years. We 
attempt to summarise here what appear to be the main 
results of these observations. Given the determination of the model parameters
by the observations presented so far, we consider how the remaining data 
might be interpreted.
\subsection{Luminosity States}
The profile
fitting of CIV fixes the inner radius of the BLR (hence the black hole mass)
and the properties of the wind. In intermediate luminosity states the narrower  CIV
and MgII profiles
are obtained by changes in the wind parameters associated with decreased
luminosity (Sect. 4.2).  At low states most of the BLR must disappear. This could
be attributed to obscuration of the inner clouds or to the suppression of cloud
formation in the BLR.
Obscuration would have to be restricted to the clouds (and not affect the central
continuum source) since there is
no correlation with, for example, X-ray absorption features. It might
therefore refer to an obscuring disc; in any case it would require a special 
line of sight, not only for NGC 4151 but presumably also for all similar 
intermediate Seyferts and for radio galaxies that lose their broad lines 
intermittently (for example, 3C390.3). 
In NGC 5548 there was no evidence for obscuring material in 
X-rays when the broad lines disappeared (Loska et al. 1993). We therefore
propose to associate the low state with a lack of inner region clouds. 
A unique feature of the model is the short timescales over which the inner
clouds are removed once injection is stopped (of order a cloud sound crossing
time, 3$\times 10^{4}r_{16}^{2}$ s). 
Thus, the time lag for the change between intermediate Seyfert types following
a transition in the luminosity state is
by the light-crossing time of the inner part of the BLR, not by structural
changes in the BLR or of some outer obscuring regions.
\subsection{ Line Widths} 
In its relatively infrequent highest luminosity state
the relative widths of the CIV, MgII and CIII] lines
are exceptional amongst AGN. At their broadest the FWHM for CIV and Mg II are between
4000 and 6000 km s$^{-1}$. The corresponding FWZI are 30 000 and 
20 000  km s$^{-1}$. Unusually for AGN, 
CIII] $\lambda$1909 is much narrower with a FWHM of 1600  km s$^{-1}$
and FWZI of 3500  km s$^{-1}$ (Clavel et al 1990). This arises from the high 
density of the inner BLR clouds ($>10^{10}$cm$^{-3}$)
which supresses the CIII]$\lambda$1909/CIV$\lambda$1549 ratio. 
Sometimes MgII shows the
characteristic double peak of CIV and sometimes the blue peak is either absent
or very weak (Ulrich et al. 1984, 1991). In the low state the CIV and MgII 
lines are much narrower, around 4000 km s$^{-1}$ HWZI, 
with suggestions of a broad base, while CIII] is
unchanged. Thus, in the low state, all three lines have similar profiles.
                                           
The model fit to
the profiles requires both a small radial extent for the BLR and
high density clouds. This results in the similarity of MgII 
and CIV line widths (FWHM) while CIII] is much narrower. 
Although this is not shown, MgII is somewhat narrower than CIV, as required, with the
difference diminishing at lower luminosity.

However, MgII does not always show the characteristic double peak of the CIV
line. Sometimes the blue peak is either absent or very weak (Ulrich et al. 
1991). We can obtain such behaviour by supressing the MgII emission from clouds
immediately leaving the disc which also gives a narrower line profile in
agreement with Ulrich et al. However, such behaviour does not appear to 
arise naturally in the model.                             

In Fig. 3 we show the effect of changing the obscuration properties of the disc
compared with the CIV line profile observations on Mar 21 and Mar 25 1991 from
Ulrich et al. (1991). We
allow the disc to become completely opaque at less than $1 \times 10^{16}$ cm 
and at less than $5 \times 10^{16}$ cm respectively, so emission from the 
cloud distribution on the far side of the disc is more completely removed. 
Equivalently, we may regard these figures as showing the effect of a 
changing albedo for a completely obscuring 
disc. We have not been able to mimic these changes by any other combinations of
alternative parameters or as variability effects (which would be included
as a radial dependence in the cloud illumination). The change               
on the red side of the blue peak (with little or no change on the
blue side) is a signature of a flared disc, the inner
clouds initially moving inwards as they leave the disc. 
                          
We show the effect on the CIV line of a reduction in the  
efficiency of cloud acceleration in fig. 4. The width of the line is reduced,
although this reduction is limited by the rotational and injection components 
of the cloud motions. Penston et al. (1981) show a sequence of CIV profiles
between Feb 1978 and Jan 1979 as the continuum varied from
the high to intermediate state. The most prominent feature is the disappearance
of the profile wings. We would therefore like to associate the reduction in cloud
acceleration with the decrease in luminosity. This cannot be the direct effect of
radiation pressure acting on the clouds which cannot accelerate clouds to
more than the internal sound speed over a sound crossing time. Since the
survival time of the inner clouds (of order $10^{5}$s) is less than the
$10^{6}$s it takes for the nuclear wind to reach the BLR it is possible 
for a reduction in the wind power to influence the cloud velocities in the
wings of the lines. Since the
cloud velocity depends on $L_{m}/Lv_{w})$ (equation (\ref{vc})) 
the line width will
decrease with $L$ if the flux of momentum in the wind decreases with $L$ more
strongly than direct proportionality. If $L_{m}/v_{w} \propto L^{2}$ then fig.
3 represents the effect of an order of magnitude change in luminosity after the
effect of a reduction in wind power has had time to propagate through the BLR..          

Further reduction of the cloud acceleration
has no effect on the profile widths since
the rotational velocity now dominates. The resulting profiles are too wide
to represent the low state observations. However, as the luminosity drops the
inner radius of the BLR will move out (equation (4)).
We obtain the further reduction in
the profile width required in the low state if we simply suppress
 the cloud formation within a radius of about
$6.5 \times 10^{16}$ cm, consistent with our estimate for $l_{r}=0.01$ of
$5 \times 10^{16}$cm. This is in agreement with Peterson (1988)
and Ulrich (1986) who show that in the lowest state virtually all of the
line emission arises from outside the high density region in the BLR. The 
remaining BLR clouds provide                        
a line of the observed FWHM (Fahey et al. 1991). Even so there is still a marked 
deficit in the core; in our model this missing flux must come from a distinct
region.  Fig. 5 shows the comparison with the data of Fahey et al.
(1991). The width of this missing emission is similar to that 
of the CIII] line and the ratio of flux in the CIV core to that in the broad
component of CIII] is normal for AGN! 
 
Our profile for CIII] from the BLR, normalised to the same peak intensity,
would be similar to CIV, in complete contradiction to
that observed. However, the high inner cloud density means that the flux in
this component of the CIII] line is merely 10 per cent of the observed flux. 
The remaining flux
consists of 70 per cent from the NLR (Ulrich et al. 1984) leaving a missing 
20 per cent which we attribute to the ILR. The computed BLR profile then
forms a shallow broad base to the observed line.

These observations clearly suggest that the bulk of the CIII]
is coming from a separate region in which the CIV/CIII] ratio is typical.
This is the intermediate line region (called BLR3 in Ulrich et al. 1986). 
It is interesting
to note that the ILR is a prediction of the model resulting from just the
fit to the profile of CIV in the high and low states. We get additional
confirmation of this from another set of low state profiles (Ulrich et al. 
1984) in which the BLR contribution is almost entirely absent and the
CIV, MgII and CIII] profiles are very similar.
We can show that the thermal disc wind gives rise to sufficient material to
account for the CIII] emission. The flux in CIV is of order $10^{11}$ erg
cm$^{-2}$ s$^{-1}$ which translates (see Table) to a luminosity in H$\beta$ 
of order $10^{7}L_{\odot}$ at 20Mpc. (We have not found a directly measured value
for the H$\beta$ flux.) Since CIII]/H$\beta$ $\sim 3 $ whereas photoionisation
models of the ILR type clouds typically give CIII]:H$\beta$ $\sim$ 1, about 1/3
of the H$\beta$ flux must be from the ILR. From standard recombination theory
we can estimate the mass in the ILR as $30
\times M_{\odot}(L(H\beta)/10^{9}L_{\odot})(10^{9} {\rm cm}^{-3}/n_{e})$. 
The average density is in the range 
$10^{8}$ cm$^{-3}$ to $3 \times 10^{7}$ cm$^{-3}$
between  $r_{c}$ and $3r_{c}$(equation (5)). 
Thus the mass required to provide all the
CIII] emission is between 1-10 $M_{\odot}$. This is close to our estimate
of the mass available (equation (30)). Thus  the disc
wind can provide sufficient material to account for the CIII] emission from the
ILR, although we cannot rule out some contribution from other sources.
Furthermore the existence of the ILR is reinforced by variability 
observations: Ulrich et al. (1984) found that CIII] does not vary on a time
scale of less than a year. Snijders (1990) has detected an underlying weak
broad variable component which could be attributed to the BLR. This implies that
the BLR, as proposed here, and the ILR cannot merge but 
that there is a gap between the outer
the BLR and the inner ILR.

It is important to note that NGC 4151 is far from unique in possessing an
ILR. IUE spectra of 3C 390.3 between 1978 and 1986 (Clavel and Wamsteker 1987)
show changes between broad high state BLR profiles and narrow low state 
profiles which we can attribute almost entirely to an ILR. Yee and Oke (1981) 
compared the luminosity of the nuclear component with that in the Balmer 
lines for 3C 390.3 and 3C 382 and found that the permitted line flux departs 
significantly from a direct proportional dependence on the nuclear continuum. 
For both objects the 1980 scans show a different behaviour from those of 
earlier epochs. Firstly, both objects showed marked changes in the Balmer 
line profiles. Secondly, between 1979 and 1980 the Balmer line fluxes of 
each had changed significantly without an accompanying change in the continuum
luminosity. (The same behaviour is exhibited by 3C 120 (Oke et al. 1980).) A 
simple time lag cannot explain completely the data for 3C 390.3 because it 
would produce much more scatter in the relation between the
line flux and nuclear continuum. Therefore Yee and Oke suggested 
a two component BLR, one component reacting on a short timescale and the other 
on a long timescale. This second component is equivalent to our ILR.

Crenshaw, Peterson and Wagner (1988) found that the decomposition of
profiles into broad and narrow components in 3C 445 leaves a 
residual component for H$\beta$ and H$\alpha$. They claim that this component
probably arises from a region kinematically distinct from the narrow
[OIII] emitting region. 
 
From the [OIII] profiles of Seyfert 1 galaxies van Groningen \& de Bryun (1989)
found that 10/12 objects possess broad wings implying densities in the region 
of 10$^{6}$ cm$^{-3}$. They call this a transition region intermediate between 
the BLR and NLR.

\subsection{Line ratios} 
The average CIV $\lambda$1549 / CIII] $\lambda$1909 
and CIV$\lambda$1549/MgII$\lambda$2798 line 
ratios are much higher than is typical
in AGN or quasars: Clavel et al. (1990) found ratios of 7.2 and 6.3
respectively. However, these values 
are highly variable and were found to correlate positively with
the continuum flux. In the December 1989 outburst Clavel et al. found CIV/CIII]
 = 11.0 and CIV/MgII = 8.0. In the wings of the lines the ratios were even
higher with CIV/CIII] $>$ 50 and CIV/MgII $\simeq 10.$ Similarly, Ulrich et al.
(1991) find CIV/CIII] $>$ 10 for $|v| > 2400$ km s$^{-1}$ and
CIV/CIII] $>20$ for $|v| > 4000$ km s$^{-1}$, in this case independent of the 
value of $f_{\lambda}(1455 \AA).$ Nevertheless, in the line cores, 
the CIV/CIII] ratio is $\sim 5$, similar to values found in typical quasar spectra.

In table I we show the observed line ratios and the model
predictions for two representative ionising continua. The Mathews and Ferland
spectrum is an average over observed AGN. However, NGC 4151 is not an average
object. In our picture this is simply a consequence of a rather special line of
sight. But, even if the continuum emission from AGN is anisotropic,  
in NGC 4151
we see just that ionising spectrum incident on the BLR clouds (as modified by
the presence of the clouds in our line of sight). 
It is possible therefore that the
ionising continuum more closely resembles the unreprocessed power law with
slope -0.9 (e.g. Nandra \& Pounds 1994). To illustrate the difference this might make we
have shown in table 1 the results from the model of a power law slope -1 for
various luminosity states. (The slope in X-rays is only weakly dependent on
luminosity (Yaqoob \& Warwick 1991).) Also shown for comparison is the photoionisation model
calculation of Ferland and Mushotsky (1982) for a homogeneous BLR and an
extrapolated spectrum corrected for X-ray absorption. It is
interesting to note the improved agreement for MgII with the power law
spectrum (although this conclusion is sensitive to any change in the
column density).

In the high state the observed low CIII]/CIV ratio  arises
from the high density clouds in the BLR. At lower luminosities the contribution
from the ILR is increasingly important and the ratios are predicted to
become closer to typical values. The low state ratios do not appear to
be given in the literature, but since the low state profiles involve the 
line cores where the high state ratios are more normal, we expect
this prediction to be borne out. 

Note that, as for other inhomogeneous models, there 
is no incompatibility between the ionization parameter deduced from line
ratios and that deduced from reverberation maps of the size of the BLR. Note
also that for the parameters of the model appropriate to NGC 4151 the clouds
are found to move only a relatively small radial distance from their point of 
origin on the disc; the mean gas density in the clouds therefore follows fairly
closely $n(r) \propto 1/r^{2}$. (This is not the case, in general, 
in systems where the normal 
injection speed is lower or where the BLR has a larger
radial extent.)

\subsection{Blue shifted `absorption'} A double peak in the CIV$\lambda$1549 
line was first noted
by Anderson and Kraft (1969). Cromwell and Weyman (1970) and Ferland and 
Mushotzsky (1982) also report a {\em variable} feature on the blue side of the
CIV line which they discuss in terms of variable absorption. 
Profiles showing this feature in a number of different
luminosity states of NGC 4151 are presented by Penston et al. (1981),
Ulrich et al. (1984) and
Stoner \& Ptak (1986). From these it is clear that the 
absorption feature varies, both in wavelength and                                          
relative depth in response to the variations in the continuum

In the model a double peak in the CIV line profile is the
result of a contribution from rotational motion together with a blue shifted 
component from the injection velocity. The corresponding red shifted peak from
clouds on the far side of the disc is obscured. The position of the trough
in the profile moves towards lower velocity as the continuum luminosity
decreases in agreement with observation. Nevertheless the depth of the trough 
in the model profile in the high luminosity state is significantly shallower
than observed. It is likely therefore that some true absorption is also
occuring in this velocity range, perhaps as a result of obscuration of the
inner broad line clouds by those at larger radii. The overall absorption line
structure is complex and we plan to return to the subject elsewhere.  

\subsection{Profile asymmetries} 
In 1986-88 all the broad lines were 
asymmetric with more flux in the red wings. 
The mid-points of the FWHMs were shifted blueward with
respect to the narrow line centres by 1000 km s$^{-1}$. The narrow lines
themselves show a small blue shift with respect to the galaxy. 
Ulrich et al. (1984) find for CIV
that the HWZI in the blue exceeds that in the red. Yet the high luminosity
profile published by Fahey et al. (1991) clearly shows excess emission and
larger width in the red wing. For Mg II the red and blue wings have similar 
widths with more emission sometimes in the red.  
Note that the red wing in CIV
may be blended with HeII $\lambda$1640 and OIII] $\lambda$1663 (Clavel et al.
1987); the MgII blue
wing may be affected by He I $\lambda$2733 or the Fe II multiplets UV62 and
63. 
Large changes in the ratio of intensities in the wings between 
campaigns and different responses of the red and blue wings to continuum
variations have been observed by Ulrich et                        
al. (1991). With no reflection or
obscuration 
from the disc the red and blue wings of the high state model profile are
symmetric, but become red asymmetric with the inclusion of a non-zero albedo.
Changes in the asymmetry are obtained in the model from changes in
the disc albedo, since this affects mainly the intensity and lag of the red component.
In the low                                                     
state the model CIV (or MgII) profile is asymmetric towards the blue i.e. the
blue wing is wider and has more flux (fig. 5).
We need to resolve the problem of blending and to include non-steady state
profiles in order to make a firm comparison with the observations. 

Ulrich et al (1991) also note a transient red component in CIV possibly
connected with the narrow so-called L$_{1}$ and L$_{2}$ 
components which are visible only in the lower luminosity states.  
The L$_{1}$ and L$_{2}$ features cannot come from our BLR. They may arise
from gas associated with the linear radio structure.
The intensities of L$_{1}$ and L$_{2}$ are not correlated with the
continuum flux, but are well correlated one with the other. The features 
have a large velocity separation but any time delay between the variations
has to be less than 4 days (Clavel et al. 1987). Thus a structure 
close to the plane of the sky is indicated. If this is the radio structure then
it is consistent with our model.

\subsection{Variability} 
Ulrich et al (1984) found typical 2-folding timescales
for CIV of 19-47 days and for Mg II of 27-72 days; the shortest two-folding
timescale over a period of about 2 years of observation was 7 days for CIV
and 20 days for MgII. The light curves of CIV and Mg II are similar to that 
of the continuum (Clavel et al. 1990) with CIV undergoing changes greater than 
60 per cent but MgII less than 50 per cent. However, the CIII] line remains 
constant. The transfer functions for MgII and CIV are similar and show zero
delay. This is contrary to Clavel et al. (1990) who observed a delay of 4 
$\pm$3 days, but any discrepancies can be explained by varying continuum pulse
lengths. 
                                                 
The peak of the line response to a continuum pulse gives a characteristic 
lag time.
For NGC 4151 there is effectively zero delay in the
{\em initial}                                                 
response of the line as a result of the presence of material in the line of
sight. For a short continuum pulse the {\em peak} 
occurs near zero delay because there is a significant 
fraction of the clouds within a sampling interval of the line
of sight. In its low state, where the inner BLR clouds are absent,
clouds in the line of sight still give an immediate response, but the
{\em peak}                                                  
response now occurs between 10-20 days, corresponding to the time at which the
equal delay surfaces contain the maximum number of clouds.

For a non-spherical BLR, such as we have in the model, this lag depends 
on the length of the continuum pulse as well as the BLR geometry. For a short
continuum pulse, duration less than or equal to the sampling time, the               
inner clouds dominate the transfer function so the peak occurs at small time
delay. For longer pulses the delay is determined by the peak of the emission
reponse at larger radii. Such                        
a dependence has been confirmed for NGC 5548 by Netzer (1991).                                

In Fig. 6 we show the model response of the CIV line to a delta continuum pulse 
from zero background for a partially obscuring disc with zero albedo. 
The lag is $\sim$ 1/2 day. Mg II and the contribution of the BLR to H$\beta$
are predicted to respond similarly. The sampling time in the only available
observations of the transfer function of NGC 4151 (Netzer 1991) is $\sim 3$ 
days. Within this resolution
the line response of CIV is indeed immediate. In the model response there is
a second peak which is the signature of the Keplerian motion of the
clouds; there is no evidence
for this in the observed data, but the resolution is almost certainly too low.

If we consider the transfer functions for the line wings and core separately
we find that changes to the bulk of the core lag those in the wings and that,
for the partially obscuring disc, 
the blue wing leads the red. The former agrees with observation; the latter 
does not. 
The 2D transfer function is difficult to construct observationally and none 
appears to be available for NGC 4151. Fig. 7
shows our prediction for the 2D transfer function for CIV with the 
same conditions as in Fig. 6 and with zero albedo for the disc. 
It is again clear that changes in the blue wing lead those in the red in
probable disagreement with observation. 

We 
remedy this discrepancy by taking a non-zero disc albedo. The transfer function  
for this reflected component is shown in fig. 8 for a reflection efficiency
of 0.3. It can be seen that incorporating reflection will give a transfer function in
which the red and blue components of the line respond with the same
lag to continuum changes.

\subsection{Profile variations} 
Gaskell (1988) found that variations 
in the red 
wings lead the blue in CIV and Mg II by a few days.
The precedence of the red wing is not confirmed by
other observers. As Gaskell points out, the exclusion of the 
{\em low state} of April 21 1980 in the cross-correlation 
function changes his result.
Clavel et al. (1990) found equal delays in the red and blue
wings of 3.2$\pm$3 days. This is consistent 
with Ulrich et al. (1991) who find delays
in the response of the wings less than their sampling time of 5 days, 
although the red and the blue wings do not always respond  to continuum 
changes in the same way. Fahey et al. (1991) found that the blue peak
of the CIV  line exhibits stronger variability than the red
peak. The emission of the red peak is also strongly correlated with
the total CIV flux, whereas the blue peak seems to be more sensitive
to variations in the CIV flux and exhibits a sharp turn-on near 
$f_{CIV} = 
10^{-11} $ erg cm$^{-2}$ s$^{-1}$.

Clavel et al. (1990) observed a 
fractional variation in the flux in
the CIV wings ( $|v| \geq$ 3000 km s$^{-1}$) of 0.34, much larger 
than that of the line core ($|v| < $3000 km s$^{-1}$), of 0.12. 
The same pattern
is observed in the MgII lines, although it is less pronounced than in the CIV 
line. Clavel (1991) found that the fractional variation
of the wing flux is as large as that of the continuum, whereas the flux in the
core only changes by $\sim$ 50\%. 
                        
Changes in the model profile which affect only the red side
of the line or the red side before the blue can only be a result of changes in
the transmittance or the reflectance of the disc. In general, if 
there are no structural changes we expect continuum variations to produce 
simultaneous changes in the red and blue wings, if we include reflection from 
the disc, and for changes on the blue side to precede those on the red if there
is no reflection. Since the unreflected line is asymmetric and the 
reflected component contributes a higher fraction
to the red than the blue the relative changes in the wings can be complex.
The fractional changes in the wings will also 
depend on the continuum pulse length, because of the 
extra path length 
for the reflected component from the outer clouds.                                

 Fig. 6 shows the transfer function 
for the CIV line as a whole. We have also constructed a diagram for the 
core of the line ($\pm 3000 $ km s$^{-1}$) only. The variation in the
core provides between
.03 and .1 of the peak throughput; the remainder is provided by the 
wings so the fractional change in the wings must exceed that in the core.

The core of CIV responds to UV variations
with a delay of 6.5$\pm$3 days, significantly longer than the response time of
the wings which is 3.2$\pm$3 days on average
(Clavel 1991). 

The 1d model transfer functions for the 
wings and line core separately show that the bulk of the core lags the 
wings in response to UV continuum changes.
                                                       
\subsection{Optical lines}
Antonucci \& Cohen (1983) observed that part of the H$\beta$ line flux
responded to continuum changes immediately, but a large part
responded only gradually. Most or all of the flux of the higher-series
Balmer lines and the $\lambda$4686 line of HeII responded immediately.
They observed no gross profile changes. Also, the features of the H$\beta$
profile are very similar in 1981 and 1975, through high and low states.
In the first half of 1984, while the nucleus was in a low state, 
De Robertis (1985) observed little, if any, broad H$\beta$ and very weak broad 
H$\alpha$ emission. 

Comparisons between H$\beta$ and CIV show that CIV is much  
broader than H$\beta$ (Osterbrock \& Koski 1976); this is especially
evident in the low luminosity state.  The change in H$\beta$  between the high
state and the low state was found to be 
qualitatively similar to that of the CIV profile
(Ptak \& Stoner 1985). The Ly$\alpha$/H$\beta$
ratio is of order 11.6 (Ferland \& Mushotzky, 1982), much higher than typical
for active galaxies; the intrinsic value could be even higher
since the blue side of the Ly$\alpha$ line is missing (Penston et al. 1979,
Ferland \& Mushotzky 1982). 
Boksenberg \& Shortridge (1975) observed that the 
HeI $\lambda \lambda$5876 and 7065 lines are twice the width of the
Balmer line wings. 

Observations by Lyutyi (1976) show that
the H$\alpha$ + [NII] emission feature varies in intensity with a 30-d
lag for the 1970-72 period and 60-70 days in 1973-76. Maoz et al. (1991)
have observed a lag in the Balmer lines of 9$\pm$2 days, still significantly 
longer than the UV line lags. 

For the dense inner BLR clouds with our high ionisation
parameter we find that the Balmer lines, are suppressed relative to
CIV (For H$\alpha$ this is clear from Table 1.) 
Thus the ILR is a significant contributor to the Balmer lines.
This is consistent with the relative
line widths, ratios and time lags. At lower luminosity the BLR contribution is 
relatively less important for the CIV line, reducing the CIV/H$\beta$ ratio
and giving lines that differ less in width.  

\subsection{Continuum}
Clavel et al. (1990) found that in the UV the typical amplitude of 
variability was the same at
1715$\AA$ and 1455$\AA$ but there did appear to be some spectral variations
which preceded the changes in flux levels by about 12 days. The ratio
of maximum to minimum flux was 3.1 for the continuum at 1455$\AA$ but
only 1.24 at 5000$\AA$. The continuum below 1900$\AA$ underwent small
but significant spectral variations. (Clavel et al. 1990). 
The X-ray luminosity in the 2-10 keV band has been found to vary between 
$\sim 2-20 \times 10^{42}$ erg s$^{-1}$ (Yaqoob et al. 1993).
The spectral variations are complex but
consistent with a picture of an intrinsically variable source, with a spectrum
that steepens on brightening. This is seen through an absorbing column that has been
modelled as cold  
($T < 10^{5}$K) gas that partially covers the source and 
varies on time scales of months to years within a range
of column densities $N_{H} \sim 5 - 15 \times 10^{22}$ cm$^{-2}$
(Barr et al. 1977, Pounds et al. 1986, Yaqoob et al. 1989) or as warm gas
where the soft X-ray opacity is reduced by partial ionization
(Warwick \& Done 1994, see also Weaver et al. 1994).
The depth of the iron K-edge implies an iron overabundance of order 
twice solar. There is very little
reddening of the optical continuum with A$_{v} \approx 0.2$ (Mushotzky et al.
1978). Therefore the X-ray absorption is not due to dust.

We have taken the spectral shape 
to be independent of luminosity state.
In fact, modest 
changes in spectrum do not appear to alter qualitatively
any of the conclusions of the paper. 

We note the possible connection between our BLR model and the X-ray 
observations. From fig. 1 we see that if the X-ray source is located
centrally then there will be some absorption from the ILR and BLR clouds
in the line of sight. The ILR is a candidate for the fixed component
of the absorbing screen (Yaqoob et al. 1989) with the BLR clouds forming
a variable partial coverer. Yaqoob et al. (1991) excluded 
the warm absorber model for this source, although Weaver et al. (1994) include a
warm absorber in a two component picture. In our model the X-ray source must be
partially covered but there could be also a scattered component and possibly
some absorption from the warm cloud debris.
The lack of reflection features from a disc 
(Maisack and Yaqoob 1991) is another indication
that the angle to the line of sight is closer to edge-on than face-on.
The complex absorption cut-off observed by Ginga is also
difficult to understand unless the line of sight is closer to being
edge-on than face-on (Yaqoob 1993). This provides some support for the 
geometry of our model.                                                  

\section{Conclusions}

We began by constructing a theoretical model for the injection and
dynamics of clouds in the broad and intermediate emission line regions. 
In this model the BLR clouds arise from the interaction between a nuclear wind
and the radiatively heated disc surface and the ILR clouds form from the
interaction of the wind with a thermally driven outflow from the disc.
When applied to NGC 4151 we found we could specify a complete set of parameters
fairly tightly from 
the profile of the CIV $\lambda$ 1549 line in the high luminosity state
of the nucleus. In particular the density of the inner clouds is high 
($\sim 10^{12}$ cm$^{-3}$) and the region is small $(10^{16}$ cm $< r 
< 10^{17}$ cm). From this and the assumption that the power in the wind drops
strongly with decreasing radiative power, 
we derived profile fits to CIV in some intermediate
luminosity states and accounted for the profile widths of other UV lines
and the Balmer lines in various states. From the low luminosity state
behaviour we deduced the existence of an extended intermediate line region
with cloud densities $\leq 10^{8}$ cm$^{-3}$. This picture agrees with that
derived from the transfer function and correlation analysis of the response
of the lines to continuum variability. Provided we include reflection from the
disc in constructing the line response then the model is also in qualitative
agreement with what is known of the transfer function (the relative
responses of red and blue wings and line core). The geometry deduced from
our model BLR also leads us to suggest an interpretation of the X-ray 
absorption data to which we shall return in more detail elsewhere.

Of other possible models for NGC 4151 pure disc motion appears to be ruled out
since variations in the line core would then {\em precede} that in the wings.
In the Mathews model of oscillating clouds (Mathews 1993) the reponse of the
core and wings should be {\em simultaneous}, which it is probably not. 
In the picture of Dyson and Perry
(1985) it is difficult to see how the red and blue wings of the line profiles
could change simultaneously in a spherical outflow. A biconical model could be
made to fit the available data for NGC 4151, but this would probably require us
to treat this galaxy as an exception since the 2D transfer function of NGC 5548
(Horne et al. 1991) appears to exclude a biconical model for that galaxy.

In our model NGC 4151 is a `special' case only in as much as we require a particular orientation
to the line of sight. We shall show elsewhere that essentially the same
model can be constructed for NGC 5548, except that the line of sight is
at greater inclination to the disc. We shall also show how  
NGC 4151, which is usually regarded as an exceptional case,            
can be fitted into a unified picture for active galaxies.

\section{Acknowledgements}
We thank C.Done and R.S.Warwick for discussions on the X-ray properties of
NGC 4151 and an anonymous referee for constructive comments.
\section*{References}

Alexander, T. \& Netzer, H., 1994, MNRAS, 270, 781\\
Anderson,K.S., Kraft, R.P., 1969, Astrophys. J., 158, 859 \\
Antonucci, R.R.J., Cohen, R.D., 1983, Astrophys. J., 271, 564 \\
Aretaxaga, I. \& Terlevich, R., 1994, MNRAS, 269, 462\\
Barr, P., White, N.E., Sanford, P.W., Ives, J.C., 1977, MNRAS 181, 43P \\
Begelman, M.C., McKee, C.F., Shields, G.A., 1983, Astrophys. J., 271, 70 \\
Blandford, R.D.\& McKee, C.F., 1982, Astrophys. J., 255, 419 \\
Blumenthal, G.R.\& Mathews, W.G., 1975, Astrophys. J., 198, 517\\ 
Boksenberg, A., Shortridge, K., 1975, MNRAS, 173, 381 \\
Capriotti, E., Foltz, C. \& Byard, P., 1980, Astrphys. J., 241, 903\\
Carroll, T. \& Kwan, J., 1985, Astrophys. J., 288, 73\\
Cassidy, I., Raine, D.J., 1993, MNRAS, 260, 385 \\
Cassidy, I., 1994, PhD Thesis, University of Leicester\\
Chandrasekhar, S., 1961, Hydradynamic and Hydromagnetic Instability, Oxford
University Press, p. 483 \\
Clavel, J., Altamore, A., Boksenberg, A., Bromage, G.E., Elvius, A., Pelat,
D., Penston, M.V., Perola, G.C., Snijders, M.A.J., Ulrich, M.-H., 1987,
Astrophys. J., 321, 251 \\
Clavel, J., Wamsteker, W., 1987, Astrophys. J., 320, L9  \\
Clavel J., Boksenberg, A., Bromage, G.E., Elvius, A., Penston, M.V., Perola,
G.C., Santos-Lleo, M., Snijders, M.A.J., Ulrich, M.-H., 1990, MNRAS, 246, 668 \\
Clavel, J., 1991, Variability in Active Galactic Nuclei, eds. H.R. Miller,
P.J. Wiita, Cambridge University Press, p. 301 \\
Collin-Souffrin, S., 1987, Astron. Astrophys., 179, 60\\
Collin-Souffrin, S., Dyson, J.E., McDowell, J.C., Perry, J.J., 1988,
MNRAS, 232, 539 \\
Crenshaw, D.M., Peterson, B.M.,  Wagner, R.M., 1988, Astron. J., 96, 1208 \\
Cromwell, R., Weyman, R., 1970, Astrophys. J., 159, L147 \\
Czerney, M. \& King, A.R., 1989, MNRAS, 241, 839\\
Emmering, R.T., Blandford, R.D., Schlosman, I., 1991, Astrophys. J., 385, 460 \\
Evans et al., 1993, \\
Fahey, R.P., Michalitsianos, A.G., Kazanas, D., 1991, Astrophys. J., 371, 136 \\
Ferland, G.J., Mushotzky, R.F., 1982, Astrophys. J., 262, 564 \\
Filippenko, A.V., 1991, Physics of Active Galactic Nuclei, eds. W.J. Duschl,
S.J. Wagner, Springer Verlag, p. 345 \\
Gaskell, C.M., 1988, Astrophys. J., 325, 114 \\
Krolik, J., 1991, Physics of Active Galactic Nuclei, eds. W.J. Duschl,
S.J. Wagner, Springer Verlag, p. 173 \\
Krolik K.J., McKee, C.E. \& Tartar, C.B., 1981, Ap.J., 249,422\\
Loska, Z., Czerny, B., Szczerba, R., 1993, MNRAS 262, L31 \\
Lyutyi, V.M., 1976, Soviet Astronomy, 21, 655 \\
Maisack, M. \& Yaqoob, T.,1991, Astron.Astrophys., 249, 25\\
Mathews, W.G. \& Doane, J.S., 1990, Astrophys. J., 352, 423 \\
Mathews, W.G., 1993, Astrophys. J., 412, L17 \\
Maoz et al., 1991, Astrophys. J., 367, 493 \\
Mobasher, B. \& Raine, D.J., 1989, MNRAS, 237, 979 \\
Mobasher, B. \& Raine, D.J., 1990, MNRAS, 244, 652\\
Mushotzky, R.F., Holt, S.S., Serlemitsos, P.J., 1978, Astrophys. J., 225, 
L115 \\
Nandra, P.   \& Pounds K.A., 1994, MNRAS, 268, 405\\ 
Netzer, H., 1991, Active Galactic Nulcei, eds. R.D. Blandford, H. Netzer,
L. Woltjer, p. 57 \\
Oke, J.B., Readhead, A.C.S., Sargent, W.C.W., 1980, PASP, 92, 758 \\
Osterbrock, D.E., Koski, A.T., 1976, MNRAS, 176, 61P \\               
Penston, M.V., Clavel, J., Snijders, M.A.J., Boksenberg, A., Fosbury, R.A.E.,
1979, MNRAS 189, 45P \\
Penston, M.V. et al.,1981, MNRAS 196, 857 \\
Penston, M.V., 1991, Variability of Active Galactic Nuclei, eds. H.R. Miller,
P.J. Wiita, Cambridge University Press, p. 343 \\         
Perez, E., Robinson, A., de la Fuente, L., 1992, MNRAS, 256, 103 \\
Perola G.C. et al., 1982, MNRAS, 200, 293\\
Perry, J.J., Dyson, J.E., 1985, MNRAS, 213, 665 \\
Peterson, B.M., 1988, PASP, 100, 18 \\
Ptak, R.L., Stoner, R., 1985, Bull. AAS, 17, 846 \\
de Robertis, M., 1985, Astrophys. J., 289, 67 \\
Robinson, A., 1995, MNRAS, 272, 647\\
Shields, G.A., 1978, Proc. Pittsburgh Conf. on BL Lac Objects, ed. A.M.Wolfe, 
p. 257 \\
Smith, M.D., Raine, D.J., 1984, MNRAS, 212, 425 \\
Smith, M.D., Raine, D.J., 1988, MNRAS, 234, 297 \\
Snijders, M.A.J., 1990, Variability of Active Galaxies, Proceedings Heidelberg,
eds. W.J. Duschl, S.J. Wagner, M. Camenzind \\
Stoner, R.E., Ptak, R., 1986, New Insights in Astrophysics, ESA SP-263, 
p. 605\\
Ulrich, M.-H., Boksenberg, A., Bromage, G.E., Clavel, J., Elvius, A., Penston,
M.V., Perola, G.C., Pettini, M., Snijders, M.A.J., Tanzi, E.G., Tarenghi, M.,
1984, MNRAS 206, 221 \\
Ulrich, M.-H.,1986, Structure and Evolution of Active Galactic
Nuclei, eds. G. Giuricin et al., Dordrecht: Reidel, p. 275 \\
Ulrich, M.-H., Boksenberg, A., Bromage, G.E., Clavel, J., Elvius, A., Penston,
M.V., Perola, G.C., Snijders, M.A.J., 1991, Astrophys. J., 382, 483 \\
van Groningen, E., de Bruyn, A.G., 1989, Astron. Astrophys., 211, 293 \\
Warwick, R.S. \& Done, C., 1994, unpublished\\
Weaver, K.A., et al. 1994, Ap.J., 423, 621\\
Yaqoob, T., Warwick, R.S., Pounds, K.A., 1989, MNRAS 236, 153  \\
Yaqoob, T. \& Warwick, R.S., 1991, MNRAS, 248, 773\\
Yaqoob,T.,Warwick, R.S., Makino, F., Otani,C., Sokoloski, J.L., Bond, I.A., \&
Yamauchi, M.,1992, MNRAS, 262, 435\\
Yee, H.K.L.\& Oke, J.B., 1981, Astrophys. J., 248, 442 \\
Zheng, W., Binette, L. \& Sulentic, J.W., 1990, Astrophys. J., 365, 115\\

\newpage
\section*{Figure Captions}
{\bf Fig. 1} Sketch of the line emission regions of NGC 4151. The BLR lies
between $r_{in}$ and $r_{out}$ in a region of the disc dominated by external
illumination where strong Kelvin-Helmholtz instabilities at the interface with
the nuclear wind, labelled $v_{w}$, lead to cloud formation and entrainment.
The ILR lies between $r_{ic}$ where Compton heating is sufficient to drive
material from the disc and $r_{iso}$ where the heating timescale exceeds the
escape time.  The disc is probably optically thick at small radii and
thin at large radii; various assumptions are explored in the text.\\
\\
{\bf Fig. 2}. The observed high state CIV profile (solid line) compared with the
model profile for 
a disc of zero albedo which partially covers clouds on the far side                
(dotted line) and for a disc with an albedo of
0.3 with clouds on the far side totally obscured (dashed line). 
The model lines have been scaled vertically to give
the best fit. Velocities on the horizontal axis are given in km s$^{-1}$ with
negative velocities to the red.\\
\\
{\bf Fig. 3.} The effect of changing disc obscuration. The solid
line is for a disc which obscures totally at distances $\leq 5 \times 10^{16}$
cm and the dashed line, on the same vertical scale, 
is for total obscuration at $\leq 1 \times 10^{16}$ cm.
The dotted line is the profile difference. From the symmetry of the model these
profiles effectively show also the effect of increasing albedo (from 0 to 1 in
the region $1\times 10^{16}$cm to $5 \times 10^{16}$cm).\\
\\
{\bf Fig. 4.} A sequence of 3 CIV model profiles, on the same vertical scale,
from high to intermediate states. The profile narrows as the
acceleration of clouds by the wind decreases corresponding to decreasing
luminosity states.\\
\\
{\bf Fig. 5} CIV profile fit for the low luminosity state. The observed
profile is shown as the solid line, the BLR contribution is the dot-dashed
line, scaled vertically to fit the bulk of the line outside the core,
and the difference is the dashed line. The difference profile is the
contribution from the ILR in the model.\\
\\
{\bf Fig. 6} One dimensional model transfer function for the CIV line (a) in a
low-luminosity state ($l_{r}=0.01$, dashed line) and (b) in a high state
($l_{r}=0.1$, solid line). The number of half day intervals is marked  
on the $x$-axis.\\
\\
{\bf Fig. 7} Two dimensional model transfer function for the CIV line (with a
partially obscuring disc and
zero albedo). The time is given in half-day intervals and the velocities in km
s$^{-1}$. \\
\\
{\bf Fig. 8} Two dimensional model transfer function for the component of the
CIV line reflected by a disc with constant albedo.
The time is given in half-day intervals and the velocities in km s$^{-1}$.
\end{document}